\documentclass[aps, prd, showpacs, floatfix, superscriptaddress, twocolumn, nofootinbib, reprint, longbibliography]{revtex4-2}
\usepackage{multirow, amsmath, hyperref, graphicx, booktabs}
\usepackage[T1]{fontenc}
\usepackage[utf8]{inputenc}
\usepackage{lmodern}
\usepackage{amsfonts}
\usepackage{array}
\usepackage{amssymb}
\usepackage{float}
\usepackage{url}
\usepackage{mathtools}
\usepackage{textcomp}
\usepackage{parskip}
\usepackage[dvipsnames]{xcolor}
\usepackage[autostyle]{csquotes}
\usepackage{orcidlink}
\usepackage[final]{microtype}
\usepackage{subcaption}
\usepackage{hyperref}
\usepackage{empheq}
\hypersetup{
colorlinks=true,
citecolor=cyan,
linkcolor=magenta,
urlcolor=NavyBlue,
allbordercolors={0 0 0},
pdfborderstyle={/S/U/W 1}
}
\newcounter{mnotecount}[section]
\newcommand{\mnotex}[1]
{\protect{\stepcounter{mnotecount}}$^{\mbox{\footnotesize
$
\bullet$\themnotecount}}$ \marginpar{
\raggedright\tiny\em
$\!\!\!\!\!\!\,\bullet$\themnotecount: #1} }

\begin{document}

\title{Ladder Symmetry: The Necessary and Sufficient Condition for Vanishing Love Numbers}

\author{Chanchal Sharma}
\email{chanchal.sharma@iitgn.ac.in}
\author{Shuvayu Roy}%
\email{shuvayu.roy@iitgn.ac.in}
\author{Sudipta Sarkar}
\email{sudiptas@iitgn.ac.in}
\affiliation{Indian Institute of Technology, Gandhinagar, Gujarat 382055, India}%


\date{\today}

\begin{abstract}

Black holes in four-dimensional, asymptotically flat general relativity have vanishing static tidal Love numbers (TLNs), a property tied to a hidden symmetry of the perturbation equations. Within the Konoplya–Rezzolla–Zhidenko (KRZ) parametrization, a subclass of spacetimes was previously shown to admit such \emph{Ladder symmetry}, which enforces the absence of static scalar TLNs. In this work, we introduce parametric deformations to such Ladder-symmetric spacetimes and analyze the resultant linear tidal response. Using the parametrized formalism for TLNs, we show that any deviation from a Ladder-symmetric background leads to non-zero static scalar TLNs. This establishes Ladder symmetry as a \emph{necessary, as well as sufficient condition} for the vanishing of static TLNs in static, spherically symmetric black holes and in rotating black holes of the KRZ class.
\end{abstract}

\maketitle


\section{Introduction}\label{sec:intro}

The study of tidal interactions in compact objects has become an important aspect of modern strong field gravity and relativistic astrophysics, especially with the advent of gravitational wave astronomy. Tidal interactions capture how compact bodies such as neutron stars and black holes respond to external gravitational fields generated by their companions. The central diagnostic quantity of this response is provided by the \emph{tidal Love numbers} (TLNs), which measure the induced multipole moments of a self-gravitating body when it is subjected to an external tidal perturbation. In physical terms, TLNs quantify the deformability of compact objects and connect the microscopic physics of dense matter with macroscopic astrophysical observables.

The concept was introduced in the Newtonian setting by Love \cite{Love:1909}, who studied the deformation of elastic bodies under tidal forces. In the relativistic context, TLNs were later extended to neutron stars \cite{Hinderer:2007mb, Damour:2009vw, Binnington:2009bb, Landry:2015cva}. This extension has profound astrophysical significance because TLNs provide a direct avenue to probe the equation of state of ultra-dense nuclear matter. Gravitational wave observations are particularly sensitive to tidal effects during the late inspiral phase of binary neutron star mergers, where the accumulation of finite-size corrections leaves observable signatures on the waveform. The landmark detection of GW170817 demonstrated this potential by delivering constraints on neutron star radii and on the nuclear equation of state \cite{Abbott:2017xzu, Abbott:2018exr, Dietrich:2017aum}.

In contrast, black holes in four-dimensional and asymptotically flat general relativity (GR) display the remarkable property that their \emph{static} TLNs vanish identically. This universality was first established for Schwarzschild black holes \cite{Damour:2009vw, Binnington:2009bb} and later extended to the rotating Kerr family \cite{Chia:2020yla, LeTiec:2020spy}. Recent studies have further clarified this picture by systematically analyzing both static and dynamical tidal Love numbers of Schwarzschild and Kerr black holes in gravitational, scalar, and electromagnetic tidal fields, showing that while static Love numbers vanish, dynamical tidal responses are generically nonzero for rotating black holes. \cite{Bhatt_2023, Bhatt_2025, Bhatt_20252}.\footnote{By contrast, for lower-dimensional black holes such as the BTZ solution, both static and dynamical scalar Love numbers are generically nonvanishing \cite{Bhatt_20242}.} The absence of tidal deformability has become a defining feature of classical black holes, motivating its use as a null test of GR with asymptotically flat boundary conditions. \footnote{The role of asymptotic boundary conditions in the vanishing of tidal Love numbers is highlighted in \cite{Nair:2024mya,Franzin:2024cah}. Also, \cite{Katagiri:2023yzm, Chakraborty:2023zed} addresses the instability of tidal Love numbers against environmental perturbations.}  Any observation of a non-vanishing TLN in connection with a black hole would constitute direct evidence for new physics beyond vacuum Einstein’s theory \cite{Cardoso:2017cfl,Pani:2015,Poisson:2014book}.

This observation naturally leads to the question of whether the vanishing of static TLNs is simply an accidental property of black holes or whether it arises from a deeper underlying principle. A strong candidate for such a principle is the so-called \emph{Ladder symmetry}, a hidden algebraic structure that appears in the perturbation equations of certain black hole spacetimes \cite{Hui:2021vcv, Sharma:2024hlz}. Owing to this symmetry, the perturbation Hamiltonian for scalar, vector, and gravitational fluctuations in both Reissner–Nordström and Kerr backgrounds can be decomposed in terms of raising and lowering operators in close analogy with the harmonic oscillator. Within this framework, the ground state solution carries zero tidal response, and repeated application of the raising operator ensures the vanishing of TLNs for all higher modes.

Building on this idea, ref. \cite{Sharma:2024hlz} carried out a systematic study of Ladder symmetries in static and stationary black hole spacetimes beyond Kerr. A particularly useful framework is provided by the parametrization introduced by Konoplya, Rezzolla, and Zhidenko (the KRZ class), which offers a systematic and theory-agnostic description of deviations from Kerr while retaining desirable separability properties \cite{Konoplya:2018arm}. The KRZ parametrization was introduced as a model-independent framework for describing black hole spacetimes in the strong-field regime, motivated by the increasing precision of gravitational-wave and electromagnetic observations \cite{Konoplya:2016jvv}. As present and forthcoming experiments probe regions ever closer to the event horizon, it becomes necessary to adopt a systematic description of possible deviations from the Kerr geometry without committing to a specific theory of gravity. The KRZ framework addresses this need by providing a unified parametrization that captures strong-gravity effects while preserving key physical properties of black hole spacetimes, such as asymptotic flatness, the existence of a regular horizon, and, in the rotating case, the separability of perturbation equations. Beyond tidal Love numbers, the KRZ parametrization has been applied to a variety of strong-field phenomena, including quasinormal modes \cite{Konoplya:2024vuj, Bolokhov:2025uxz}, geodesic motion, and accretion-disk observables \cite{He:2025rjq, Olmo:2025ctf, Wang:2025gbj}, as well as black hole shadows \cite{Uniyal:2025uvc} and lensing properties \cite{Feleppa:2025ejh}, and has therefore been widely used in phenomenological tests of the Kerr-like geometry. By analyzing scalar perturbations in these geometries, \cite{Sharma:2024hlz} identified a subclass of KRZ spacetimes that admit Ladder symmetry and consequently established that scalar TLNs vanish for the entire subclass. This finding demonstrated a clear correspondence between Ladder symmetry and the absence of tidal deformability in a wide family of black hole geometries.

However, an important complementary question remained open. Is Ladder symmetry not only sufficient but also necessary for the vanishing of TLNs? In other words, could there exist black hole spacetimes that do not admit Ladder symmetry but nevertheless exhibit vanishing TLNs, at least in the scalar sector?

The resolution of this question carries significant implications. On one hand, it clarifies the geometric and algebraic origin of the no Love property in GR. On the other hand, it addresses whether departures from Ladder symmetry—arising from modifications of gravity, quantum corrections, or astrophysical environments—necessarily lead to observable tidal responses. Establishing both necessity and sufficiency would elevate Ladder symmetry from a mathematical curiosity to a fundamental physical principle governing the tidal properties of black holes.

The present work aims to settle this question. We show that for static spherically symmetric black holes as well as rotating black holes within the KRZ class, Ladder symmetry is both a necessary and a sufficient condition for the vanishing of static TLNs. To demonstrate this, we employ the parametrized TLN framework developed in \cite{Cardoso:2018ptl, Katagiri:2023umb} in combination with the most general Ladder-symmetric spacetimes classified in \cite{Sharma:2024hlz}. By perturbing away from Ladder-symmetric backgrounds and analyzing the induced modifications to the effective perturbation potential, we establish that any departure from Ladder symmetry generically produces non-vanishing TLNs. This completes the logical equivalence and firmly identifies Ladder symmetry as the fundamental principle behind the no Love property of classical black holes.

\section{Setup}\label{sec:setup}
In this section, we review the earlier results that provide the foundation for our present framework and outline the strategy that enables us to address all the relevant cases of interest. In Sec.~\ref{subsec:parform}, we begin with a brief summary of the parametrized formalism for the computation of tidal Love numbers. This includes the necessary definitions, conventions, and notational choices, which will serve as the common language throughout the remainder of this work. 

Subsequently, in Sec.~\ref{subsec:oldsharmaghosh}, we revisit the results of Ref.~\cite{Sharma:2024hlz}. These results establish the most general forms of Ladder-symmetric spacetimes belonging to the Konoplya--Rezzolla--Zhidenko (KRZ) parametrization, both in the context of static, spherically symmetric black hole geometries and for stationary, rotating, axisymmetric black hole solutions. For the present analysis, we shall also demonstrate that it is sufficient to formulate the problem directly at the level of the perturbation (wave) equation, rather than starting from the full metric description. This observation significantly simplifies our calculations while retaining complete generality. 

Finally, in Sec.~\ref{subsec:strategy}, we present the strategy underlying our approach. Using the Ladder-symmetric metrics described in Sec.~\ref{subsec:oldsharmaghosh} as the background spacetimes, we will systematically quantify deviations from them within the parametrized framework introduced in Sec.~\ref{subsec:parform}. \footnote{In the static case, both the ladder-symmetric backgrounds considered here and the metrics obtained by perturbing away from ladder symmetry fall within the general class of static, spherically symmetric spacetimes described by the Rezzolla–Zhidenko parametrization \cite{PhysRevD.90.084009}. Moreover, when the KRZ framework is restricted to the non-spinning limit and separability of the Klein–Gordon equation is imposed, it reduces to the static Rezzolla–Zhidenko form up to trivial gauge choices.} This will provide a consistent and unified setting for our subsequent analysis of TLNs across different classes of black hole geometries.

\subsection{The Parametrized Formalism of calculating TLNs: A Brief Review}\label{subsec:parform}

In \cite{Cardoso:2019mqo, McManus:2019ulj, Katagiri:2023umb}, the authors develop a parametrized formalism to compute various quantities of interest on spacetimes perturbatively deviated from known background solutions in a theory-agnostic manner. In \cite{Cardoso:2019mqo,  McManus:2019ulj}, this formalism was employed for QNMs, while in \cite{Katagiri:2023umb}, it was used to derive tidal response in various scenarios. In \cite{Kobayashi:2025swn}, the authors used the same formalism to derive the tidal dissipation numbers for non-rotating black holes.

Here, we review the results of \cite{Katagiri:2023umb} and discuss in Sec. \ref{subsec:strategy} how we adapt them to our analysis. The key assumptions underlying their approach are the following: (1) the background is spherically symmetric; (2) deviations from the background metric are small; and (3) no coupling exists between different perturbation sectors. Within this setup, the standard Zerilli/Regge–Wheeler and spin-$s$ perturbation equations are deformed by introducing small, linear power-law corrections to the effective potentials. These corrections, parametrized by coefficients, $\alpha_j^{\pm/s}$, capture the possible deviations from the background geometry. 

Consider a static scalar ($\omega = 0,~s=0$) field $\Xi\,(r,\theta,\varphi) = \frac{\chi_\ell(r)}{r} Y_{\ell m}(\theta, \varphi)$, where $Y_{\ell m}(\theta,\varphi)$ are the spherical harmonics of order $(\ell,m)$. The master equation governing these scalar perturbations on the Schwarzschild background is given by the radial equation of motion of $\chi_\ell(r)$ transformed into the Fourier domain,
\begin{equation}\label{eq:masterschwarz}
    f\frac{d}{dr}\left[f\frac{d\chi_{\ell}}{dr}\right] - \left( f~V_{\ell}\right)\chi_{\ell} = 0~,
\end{equation}
with the effective potential
\begin{equation}\label{schw.potential}
    V_{\ell} = \frac{\ell(\ell+1)}{r^2} + \frac{\partial_r f}{r},~~ f = \left(1 - \frac{r_+}{r} \right),
\end{equation}
where $r_+$ is the Schwarzschild radius of the black hole.
Then, imposing regularity at the black hole horizon, and appropriate normalization far away from the horizon (i.e., $r>>r_+$), the asymptotic solution for $\chi_\ell(r)$ at large $r$ is given by
\begin{equation}
    \begin{split}
        \chi_{\ell}|_{r>>r_+} \propto & \left(\frac{r}{r_+}\right)^{\ell+1} \left[1 + \mathcal{O}\left(\frac{r_+}{r}\right)\right] \\
        &+ \kappa_{\ell} \left(\frac{r_+}{r}\right)^{\ell} \left[1 + \mathcal{O}\left(\frac{r_+}{r}\right)\right]~,
    \end{split}
\end{equation}
where $\kappa_{\ell}$ can be read off as the static TLN for scalar perturbations. Note that the above expression solely contains a power series of $r$. In more generic systems, the perturbations at large distances may include logarithmic terms besides these power series, as
\begin{equation}
    \begin{split}
        &\chi_{\ell}|_{r>>r_+} \propto  \left(\frac{r}{r_+}\right)^{\ell+1} \left[1 + \mathcal{O}\left(\frac{r_+}{r}\right)\right] \\
        &+ K_{\ell} \left[\ln\left(\frac{r}{r_+}\right)+ \mathcal{O}\left(\frac{r_+}{r}\right)^0 \right] \times\left(\frac{r_+}{r}\right)^{\ell} \left[1 + \mathcal{O}\left(\frac{r_+}{r}\right)\right]~.
    \end{split}
\end{equation}
In such cases, the prefactor of the linear response term $(r_+/r)^{\ell}$ includes a logarithmic function dependent on $r$, and the constant numbers $K_{\ell}$'s are regarded as the fundamental parameters of a given system, which are independent of the scale measured, and then evaluated as ``Running TLNs".
For Schwarzschild black holes, it is well-known that the horizon regularity condition enforces its static TLNs to vanish for all kinds of perturbations. 

Any modifications to the Schwarzschild background, under the assumptions of the parametrized formalism, change the perturbation equation (\ref{eq:masterschwarz}) to the form,
\begin{equation}\label{eq:parschwarz}
    f\frac{d}{dr}\left[f\frac{d\chi_{\ell}}{dr}\right] - \left\{ f \left(V_{\ell} + \delta V_{\ell}\right)\right\}\chi_{\ell} = 0~.
\end{equation}
Note here, $f=\left(1-\frac{r_+}{r}\right)$, with $r_+$ as the horizon radius perturbatively connected to Schwarzschild radius and $\delta V_{\ell}$ is the sum of small linearized power-law corrections to the potential in eq. (\ref{schw.potential}) parametrized in terms of $\alpha$'s as
\begin{equation}
    \delta V_{\ell} = \frac{1}{r_+^2} \sum_{j=3}^\infty \alpha_j \left(\frac{r_+}{r}\right)^j~.
\end{equation}
Here, $j=0, 1, 2$ corrections are excluded from the analyses as these would be unboundedly large, and it is assumed that $|\alpha_j|<< r_+^2 |V_{\ell}(r \to r_+)|$ that suffices to ensure the smallness of all the other corrections. The linearity of the problem allows us to extract the physical tidal responses order by order in the small parameters $\alpha_j$'s by defining a ``basis" set for TLNs. We can solve eq. (\ref{eq:parschwarz}) perturbatively by expanding $\chi_\ell$ up to the first order in $\alpha_j$ as
\begin{equation}
    \chi_{\ell} = \chi_{\ell}^{(0)} + \sum_{j=3}^\infty \alpha_j \chi_{\ell,\,j}^{(1)}~.
\end{equation}
Here, $\chi_{\ell}^{(0)}$ corresponds to the solution with the Schwarzschild background, and $\chi_{\ell,\,j}^{(1)}$ is the solution on the background that introduces a correction proportional to $\alpha_j$. The GR contribution to the TLNs is zero, and they are corrected by small linear corrections in this framework, as
\begin{equation}
    \kappa_{\ell} = 0 + \sum_{j=3}^\infty \alpha_j e_{\ell,\,j}~,
\end{equation}
with $e_{\ell,\,j}$'s as the ``basis'' set for TLNs, as termed in \cite{Katagiri:2023umb}. Similarly, we can introduce a ``basis'' set $d_{\ell,\,j}$'s for the coefficient of running TLNs as 
\begin{equation}
    K_{\ell} = 0 + \sum_{j=3}^\infty \alpha_j d_{\ell,\,j}~.
\end{equation}
The basis is determined by the TLNs that are read off from the asymptotic behavior of the first-order horizon-regular solution, $\chi_{\ell,\,j}^{(1)}$ at large distances. Thus, this formalism explicitly relates the different sets of TLNs to their underlying perturbative equations, while staying agnostic about the theory that gives rise to the perturbations.

\subsection{Background Spacetimes with Ladder Symmetry}\label{subsec:oldsharmaghosh}

In \cite{Sharma:2024hlz}, the authors employ the presence of Ladder symmetries to impose constraints on the black hole spacetime metric. Here, we revisit the Ladder-symmetric solutions obtained therein, focusing on their role as background spacetimes within the parametrized framework. Furthermore, we will show that for the analyses and results presented in the following sections, it is sufficient to adopt the wave equation as the starting point, rather than the explicit form of the metric.

The most general form of the metric for a static, spherically symmetric four-dimensional spacetime where Ladder symmetry exists can be written as \cite{Sharma:2024hlz}
\begin{equation}
\begin{split}
    &\mathrm{d}s^2 = -\frac{\Delta_b(r)}{h(r)}\,\mathrm{d}t^2 + \frac{h(r)}{\Delta_b(r)}\,\mathrm{d}r^2 + h(r)\,\mathrm{d}\Omega^2_{(2)}, \\
    &\Delta_b(r) = r^2 - c_2 ~r ~+ c_3~.
\end{split}
\end{equation}
The function $h(r)$ can be left arbitrary, except for those which would produce a singularity in our region of interest, i.e., between the outer horizon and the asymptotic infinity. The radial part of the wave equation for a static scalar field $\Xi(r,\theta,\varphi) = \chi_\ell(r) Y_{\ell m}(\theta, \varphi)$ on this background would be given by 
\begin{equation}\label{eq:staticwave0}
    -\Delta_b(r) \Bigg[ \partial_r \Big\{\Delta_b(r) \partial_r\Big\} - \ell (\ell+1) \Bigg] \chi_\ell(r) =0~.
\end{equation}
Note that the function $h(r)$ does not appear in the wave equation. Therefore, the choice of $h(r)$ has no consequences on the existence of Ladder symmetry in the solution. For convenience in later calculations, we can cast the function $\Delta_b(r)$ as
\begin{equation}\label{eq:staticdeltab}
    \begin{split}
    \Delta_b(r) &= r^2 - c_2 r+ c_3 = r^2 \left(1 - \frac{c_2}{r} + \frac{c_3}{r^2}\right) \\
    &= r^2 \left(1 - \frac{R_+}{r} \right) \left(1 - \frac{R_-}{r} \right) \\
    &\equiv r^2~ f_{sph}(r)~.
    \end{split}
\end{equation}
where, the constants $R_+$ and $R_-$ are the two roots of the equation $\Delta_b(r) = 0$. Since we are interested in the most general black hole solutions in such spacetimes and the radii of the outer and the inner horizons are given by the roots of the equation $\Delta_b(r)=0$, we will assume $R_+$ and $R_-$ to be the outer and the inner horizons, respectively, and thus, positive real numbers satisfying $R_+ \geq R_-$. Hence, we will assume $c_2$ and $c_3$ to have such values that allow the conditions mentioned above to be satisfied by $R_+$ and $R_-$.

Using the form of eq. (\ref{eq:staticdeltab}) in eq. (\ref{eq:staticwave0}) and redefining $\chi_\ell(r) = \Phi_\ell(r)/r$, the wave equation becomes
\begin{equation}\label{eq:staticwave}
    \begin{split}
        f_{sph}(r)\Bigg[&\partial_r \Big\{f_{sph}(r) \partial_r ~\Phi_\ell(r)\Big\} -  V_{sph}(f_{sph}(r),r) \Phi_\ell(r) \Bigg] = 0~,
    \end{split}
\end{equation}
where, $$V_{sph}(f_{sph}(r),r) = \frac{\ell (\ell+1)}{r^2} + \frac{\partial_r f_{sph}}{r}~.$$

In asymptotically flat four-dimensional spacetimes, the most general class of stationary axisymmetric rotating black hole solutions where the scalar wave equation splits into radial and angular parts in the Boyer-Lindquist-type coordinates is denoted by the Konoplya-Rezzolla-Zhidenko (KRZ) class of metrics \cite{Konoplya:2016jvv, Konoplya:2018arm}. The KRZ class includes well-known rotating black hole solutions like the Kerr, Kerr-Newman, and Kerr-Sen solutions. 

The most general Ladder-symmetric form of the KRZ metric can be expressed as \cite{Sharma:2024hlz}
\begin{equation}\label{eq:metricLaddSymKRZ}
    \begin{split}
        &\mathrm{d}s^2 = -\mathrm{d}t^2 \frac{\Delta - a^2 \sin^2 \theta}{r^2 R_\Sigma + a^2 \cos^2 \theta} + \mathrm{d} r^2~ \frac{r^2 R_\Sigma+a^2 \cos^2 \theta}{\Delta} \\
        &- \mathrm{d}t ~\mathrm{d}\phi~ \frac{2a~(a^2-\Delta+r^2 R_\Sigma) \sin^2 \theta}{r^2 R_\Sigma+a^2 \cos^2 \theta} \\
        &+  \mathrm{d}\theta^2~ (r^2 R_\Sigma + a^2 \cos^2 \theta)  \\
        &+ \sin^2 \theta \mathrm{d} \varphi^2~ \frac{a^4 - a^2 (\Delta~\sin^2 \theta - 2 r^2 R_\Sigma) + r^4 R_\Sigma^2}{r^2 R_\Sigma + a^2 \cos^2 \theta} \\
        &R_\Sigma(r) = R_\Sigma^{(0)} + \frac{R_\Sigma^{(-1)}}{r} + \frac{R_\Sigma^{(-2)}}{r^2} ~\\
        &R_M(r) = R_M^{(1)}r + R_M^{(0)} + \frac{R_M^{(-1)}}{r}
    \end{split}
\end{equation}
and $\Delta(r), R_\Sigma(r),$ and $R_M(r)$ are related by the condition of the existence of Ladder symmetry as
\begin{equation}
    \begin{split}
        \Delta(r) &= r^2 +A~r+ B \\
        R_\Sigma^{(0)} &- R_M^{(1)} = 1 ,~\\
        R_\Sigma^{(-1)} &- R_M^{(0)} = A ,~\\
        R_\Sigma^{(-2)} &- R_M^{(-1)} + a^2 = B~~
    \end{split}
\end{equation}
where, the constants $A$ and $B$ in $\Delta(r)$ are arbitrary numbers. Again, as before, we are interested in black hole solutions. Therefore, we will assume that $A$ and $B$ are real numbers such that they allow two positive real roots of the equation $\Delta(r) = 0$. We will again denote these two roots by $R_+$ and $R_-$. In appendix \ref{app:LaddSymKRZ}, we discuss how the possibly infinite series in $R_\Sigma$ and $R_M$ are constrained to their truncated forms in eq. (\ref{eq:metricLaddSymKRZ}) due to Ladder symmetry.

The wave equation for a static scalar field $\Xi(r,\theta,\varphi)= \chi_\ell(r) Y_{\ell m}(\theta,\varphi)$ on this background can be written as
\begin{equation}
    \begin{split}
        -\Delta(r) \Bigg[ \partial_r \Big\{\Delta(r) \partial_r \Big\}  - \ell(\ell+1) + \frac{a^2 m^2}{\Delta(r)} \Bigg] \chi_\ell(r) &= 0~.
    \end{split}
\end{equation}
Again, we can express $\Delta(r)$ here as 
\begin{equation}
\begin{split}
    \Delta(r) &= r^2 ~f_{KRZ}(r),~\\
    f_{KRZ}(r) &= 1 + \frac{A}{r} + \frac{B}{r^2} = \left(1- \frac{R_+}{r}\right) \left(1- \frac{R_-}{r}\right) 
\end{split}
\end{equation}
and recast the wave equation in the form
\begin{equation}\label{eq:KRZwave}
    \begin{split}
        f_{KRZ}(r) \Bigg[&\partial_r \Big\{f_{KRZ}(r) \partial_r\Phi_\ell(r)\Big\}\\
        &-  V_{KRZ}(f_{KRZ}(r),r)\Phi_\ell(r) \Bigg] = 0~,
    \end{split}
\end{equation}
where, $$V_{KRZ}(f_{KRZ}(r),r) = \frac{\ell (\ell+1)}{r^2} + \frac{\partial_r f_{KRZ}}{r} - \frac{a^2 m^2}{r^4 f_{KRZ}}.$$

From the forms of the wave equations given in eq. (\ref{eq:staticwave}) and eq. (\ref{eq:KRZwave}), we see that it is possible to  cast both the wave equations into the schematic form
\begin{equation}\label{eq:waveeqncommonform}
    f_0(r) \Bigg[ \partial_r \Big\{ f_0(r) \partial_r~\Phi_\ell(r) \Big\} - V_0(f_0(r),r) \Phi_\ell(r) \Bigg] = 0~,
\end{equation}
where $f_0(r)$ and $V_0(f_0(r),r)$ take their appropriate values corresponding to either case.
For both cases, $f_0(r)$ can be expressed as
\begin{equation}\label{eq:genf0}
    f_0(r) = \left( 1 - \frac{R_+}{r} \right) \left( 1 - \frac{R_-}{r} \right)~.
\end{equation}

Furthermore, $V_0(f_0(r),r)$ depends only on $f_0(r)$ and the numbers $\ell$ and $m$. Therefore, for both classes of spacetimes, we see that all the details of the underlying metric relevant to the wave equation enter through the function $f_0(r)$. Hence, since Ladder symmetry is a property of the solutions to the wave equation itself, we can infer that any deviation from Ladder symmetry would be captured at this level through the form of $f_0(r)$. 

Now, depending on the choice of the residual gauge-fixing in the KRZ class of metrics, one can end up with different ways of quantifying the deviation from $f_0(r)$ in the metric. As shown in Sec. \ref{sec:TLNsphsym}, one such possible choice leads to having two different independent functions $f_t(r) = -g_{tt}$ and $f_r(r) = g_{rr}^{-1}$ in the wave equation. This way of encoding the deviations from a background metric can also be found in the EFT-corrected metrics studied in \cite{Cardoso:2018ptl} and \cite{Barbosa:2025uau}. Without loss of generality, a different choice of the gauge studied in \cite{Sharma:2024hlz} can be implemented that leads to having only one independent function $\Delta(r)$ in the wave equation. In the following two sections \ref{sec:TLNsphsym} and \ref{sec:TLNrotating}, we will work with the choice used in \cite{Sharma:2024hlz}, such that all deviations from the background will be expressed as subleading terms in $\Delta(r)$. We will treat the metric and the resultant wave equation as our starting point and demonstrate how the parametrized formalism can be implemented in both these cases with two horizons in the background geometry.

\subsection{Strategy and Algorithm}\label{subsec:strategy}

We begin with the assumption that the equation of motion for a scalar field is separable into a radial and an angular part. Consequently, even though we deviate perturbatively from the Ladder-symmetric class of KRZ metrics, we remain within the broader class of KRZ metrics. Our aim in this work is to find whether there exists a metric in this class of solutions where Ladder symmetry doesn't exist, but static scalar TLNs vanish. 

We start with the metric of a generic stationary, rotating black hole belonging to the KRZ class \cite{Konoplya:2018arm}
\begin{equation}\label{eq:genKRZmet}
    \begin{split}
        &\mathrm{d}s^2 = - \frac{N^2 - W^2 (1-y^2)}{K^2} \mathrm{d}t^2 - 2 W~r(1-y^2) \mathrm{d}t~\mathrm{d}\varphi \\
        &+ K^2r^2 (1-y^2) \mathrm{d}\varphi^2 + \Sigma \left( \frac{R_B^2}{N^2} \mathrm{d}r^2 + r^2 \frac{\mathrm{d}y^2}{1-y^2}  \right) \\
        &W = \frac{a~R_M}{r^2 \Sigma} ,~~
        N^2 \equiv R_N = R_\Sigma - \frac{R_M}{r} + \frac{a^2}{r^2} \\
        &K^2 = \frac{1}{\Sigma} \left( R_\Sigma^2 + R_\Sigma \frac{a^2}{r^2} + \frac{a^2 y^2}{r^2} N^2 + \frac{a^2 R_M}{r^3} \right)~,
    \end{split}
\end{equation}
where, $y=\cos \theta$, and $R_M, R_\Sigma$, and $R_N$ are all functions of $r$.
The static limit of this metric can be taken by setting $a=0$ to get
\begin{equation}\label{eq:genKRZstat}
    \begin{split}
        &\mathrm{d}s^2 = - \frac{R_\Sigma - \frac{R_M}{r}}{R_\Sigma} \mathrm{d}t^2 + \frac{R_\Sigma}{R_\Sigma - \frac{R_M}{r}} R_B^2 ~\mathrm{d}r^2 + r^2 R_\Sigma~ \mathrm{d}\Omega_{(2)}^2~.
    \end{split}
\end{equation}
The Klein-Gordon equation for a massless scalar field $\Xi(r,\theta,\phi) = Y_{\ell m}(\theta,\varphi) \Phi_\ell(r)/r$ on the background eq. \ref{eq:genKRZmet} is given by
\begin{equation}\label{eq:genwaveKRZ}
\begin{split}
    &\frac{R_N}{R_B} \left[  \partial_r \left( \frac{R_N}{R_B} \partial_r \Phi_\ell \right) - \mathcal{V}(r) \Phi_\ell\right] = 0 \\
    &\mathcal{V}(r) = \left( \frac{\ell(\ell+1)}{r^2} R_B + \frac{1}{r} \partial_r \left( \frac{R_N}{R_B} \right) - \frac{a^2 m^2 R_B}{r^4 R_N} \right)~.
\end{split}
\end{equation}
Note that for axisymmetric perturbations, i.e., $m=0$, the last term drops off, and with an appropriate residual gauge-fixing, the wave equation takes the same symbolic form for the static spherically symmetric, as well as the stationary rotating spacetime.

The next assumption in our calculations is that the corrections to the Ladder-symmetric background metric are perturbative in sub-leading powers of $r^{-1}$. The resultant corrections to the effective potential of the wave equation are then admissible in sub-leading powers of $r^{-1}$, and are expressed in a parametrized form. It can be checked from eq. (\ref{eq:genwaveKRZ}) that for perturbations with $\ell=0$, there is no tidal response. Following this, we show that in generic cases with $\ell \geq 1$, there exists a logarithmic-running contribution to the static scalar TLNs. It can be shown that, for corrections to the effective potential with only a finite number of terms in powers of $r^{-1}$, there exists a minimum value of $\ell$, beyond which the static scalar TLNs can receive non-zero contributions only with a log-running behavior. Now, for a particular value of $\ell$, since at $r >> r_+$, the behaviors of the $\log (r/r_+)$ and the $(r_+/r)^\ell$ terms are very different, and it is impossible for the non-running contributions to cancel out the log-running contributions completely. Hence, to have vanishing static scalar TLNs for all $\ell \geq  1$, the logarithmic-running contributions and the non-running contributions must cancel out separately for each value of $\ell$. The running parts canceling out provides us with an infinite number of coupled constraint equations, while the non-running parts supply a finite number of the same. Thus, we end up with an infinite number of constraint equations that the parameters involved in the metric corrections must satisfy. 

To establish that the existence of Ladder symmetry is a necessary condition for static scalar TLNs to vanish for all values of $\ell$, it is sufficient to show that on a generic non-Ladder-symmetric metric, there always exists some $\ell$ value for which the TLN doesn't vanish. We will derive the parametrized corrections to the effective potential and argue that trying to solve these infinite constraint equations gets us back to the Ladder-symmetric background solution, thus making it impossible to have vanishing TLNs for all $\ell$ values in a metric that is perturbatively deviated from a Ladder-symmetric background metric.

Note that, if one were to show that there can exist metrics where TLNs vanish without the existence of Ladder-symmetry, then one would have to show that, besides the log-running contributions, the non-running contributions in the TLNs must vanish as well. Therefore, for our purpose, we could utilize these non-running contributions as well. But we will not do so for the following reasons. Firstly, in cases where the parametrized corrections to the effective potential are finite in number, there exists a maximum value of $\ell$ beyond which there will be no non-running contributions in the TLNs. Thus, even if we consider these contributions, in most cases, we will only have a finite number of such constraints on top of the infinite number of constraints obtained from the running contributions alone. Furthermore, as discussed in \cite{Katagiri:2023umb}, when the log-running contributions in the TLNs vanish identically, there can be ambiguities in reading off the pure tidal response terms from the resultant expressions, which are not resolved in the parametrized formalism. Using only the running TLNs allows us to bypass this subtlety of non-running TLNs. Therefore, we will employ only the running contributions to the TLNs to derive our main results.

\section{Parametrized Love numbers for Static Spherically Symmetric Metrics}\label{sec:TLNsphsym}

In this section, we derive the parametrized corrections to the effective potential in the massless Klein-Gordon equation on a spacetime that is perturbatively away from a background Ladder-symmetric metric. Thereafter, we show that demanding the TLNs to be zero for all values of $\ell$ gets us back to the Ladder-symmetric background spacetime. 

We start by considering a generic static, spherically symmetric, asymptotically flat metric in four dimensions, given by
\begin{equation}\label{eq:genflat}
    \mathrm{d}s^2 = - f_t (r) \frac{r^2}{h(r)} \mathrm{d}t^2 + \frac{1}{f_r(r)} \frac{h(r)}{r^2} \mathrm{d}r^2 + h(r) \mathrm{d}\Omega_{(2)}^2~.
\end{equation}
This metric is given by the following choices in eq. (\ref{eq:genKRZstat})
\begin{equation}
    R_B(r) = \sqrt{\frac{f_t(r)}{f_r(r)}},~R_N(r) = f_t(r),~ R_\Sigma(r) = \frac{h(r)}{r^2}~.
\end{equation}
We have still not fixed the residual gauge freedom in eq. (\ref{eq:genKRZstat}), and this would be a good point to shed light on some of the possible choices. The analysis of \cite{Cardoso:2018ptl} corresponds to choosing $R_\Sigma =1$. For the general rotating KRZ solution, this choice corresponds to the Boyer-Lindquist coordinates for the Kerr metric. An alternative choice of this residual gauge is to fix $R_B=1$, which corresponds to Schwarzschild-like coordinates in the $a\to 0$ limit. From eq. \ref{eq:KRZwave}, it is clear that setting either $a=0$ or $m=0$ in $V_{KRZ}$ yields equations that are schematically identical (although, the $m=0$ case still has an implicit dependence on $a$ through $R_+$ and $R_-$ in $f_{KRZ}$). This similarity implies that $m=0$ perturbations on a rotating KRZ background can be treated on the same footing as those in the $a\to 0$ (non-rotating) limit. Hence, in the subsequent calculations presented here, we use the $R_B=1$ gauge choice, which is consistent for both the static spherically symmetric and the stationary rotating KRZ metrics.

This keeps $h(r)$ as a free function of $r$. However, since it does not appear anywhere in the wave equation, this doesn't lead to any further non-trivialities. Thus, under this gauge-choice, eq. (\ref{eq:genflat}) becomes
\begin{equation}
    \mathrm{d}s^2 = - f (r) \frac{r^2}{h(r)} \mathrm{d}t^2 + \frac{1}{f(r)} \frac{h(r)}{r^2} \mathrm{d}r^2 + h(r) \mathrm{d}\Omega_{(2)}^2~~.
\end{equation}

The function $f(r)$ is given by
\begin{equation}\label{eq:statbggeneral}
    \begin{split}
        f(r) &= \sum_{n=0}^N \frac{B_{(n)}}{r^n} \\
        B_{(0)} & = 1,~~
        B_{(1)} =  -2M,~~
        B_{(2)}  = Q^2 \\
        B_{(n)} &= \epsilon~\beta_{(n)},~~  \forall ~3 \leq n \leq N~,
    \end{split}
\end{equation}
where $\epsilon$ is a small parameter to quantify the deviation from the Ladder-symmetric background metric. This can be seen by taking the $\epsilon=0$ limit of eq. (\ref{eq:statbggeneral}), which reproduces the static, spherically symmetric Ladder-symmetric background metric. 

At $\epsilon=0$, the roots of $f(r)=0$ are given by
\begin{equation}
    R_+ = M + \sqrt{M^2 - Q^2},~~~ R_- = M - \sqrt{M^2 - Q^2}~.
\end{equation}
When $\epsilon$ is turned on, the roots of $f(r)=0$ can be classified into two different categories. 
We term the first of these as the ``principal roots'', which are perturbatively deviated from the roots $R_\pm$. These can be expressed as
\begin{equation}
    \begin{split}
        r_+ &= R_+ ~+ \epsilon \sum_{n=3}^N \beta_{(n)} \rho_{+}^{(n)}\\
        r_- &= R_- ~+ \epsilon \sum_{n=3}^N \beta_{(n)} \rho_{-}^{(n)}~,
    \end{split}
\end{equation}
where, $\rho_{\pm}^{(n)}$ are $\mathcal{O}(1)$ numbers.

The second category of roots comprises the $(N-2)$ number of roots of $f(r)$. We name them as ``perturbative roots'' because they exist only when $\epsilon \neq 0$. In terms of $\mathcal{O}(1)$ numbers $\rho^{(n,n_1,n_2)}$, these can be generally expressed as
\begin{equation}
    \begin{split}
        r_{(n)} &= \sum_{n_1 =3}^N \sum_{n_2 = 1}^{n_1 - 2} (\epsilon \,  \beta_{(n_1)})^{\frac{n_2}{n_1 -2}} \rho^{(n,n_1,n_2)}~.\\
    \end{split}
\end{equation}
Further, defining
\begin{equation}
    \bar f(r) = \left(1 - \frac{r_+}{r} \right)\left(1 - \frac{r_-}{r} \right)~,
\end{equation}
we can express $f(r)$ as
\begin{equation}
    \begin{split}
        f(r) &= \bar f(r) \prod_{n=3}^N \left(1 - \frac{r_{(n)}}{r} \right)~.
    \end{split}
\end{equation}
Expanding these products in terms of $B_{(n)}$'s as discussed in appendices \ref{app:POStoSOP} and \ref{app:moreonPOStoSOP}, and defining
\begin{equation}
    H_{n,m} \equiv \sum_{p=0}^{n-m} r_+^p r_-^{n-m-p} = \frac{r_+^{n-m+1} - r_-^{n-m+1}}{r_+ - r_-}~,
\end{equation}
we get
\begin{equation}
    \begin{split}
        f(r) &= \bar f(r) \left[ 1 + \sum_{n=1}^{N-2} \frac{(-1)^n}{r^n} \sum_{m=0}^{n} H_{n,m} B_{(m)}  \right]~.
    \end{split}
\end{equation}
\begin{widetext}
The resultant wave equation for a massless scalar field on this background is then given by
\begin{equation}\label{eq:KGEstat}
    \begin{split}
        &f \left[ \partial_r (f~\partial_r \chi_\ell) - \bar V \chi_\ell \right] =0 \\
        &\bar V(r) = \frac{\ell(\ell+1)}{r^2} + \frac{1}{r}\partial_r (f)~.
    \end{split}
\end{equation}
In terms of $\bar f(r)$, we define 
\begin{equation}
\begin{split}
    f(r) &= \bar f(r) \left[ \prod_{n=3}^N \left( 1 - \frac{r_{(n)}}{r} \right)  \right]^{\frac{1}{2}}  \\
    &\equiv \bar f(r) (1 + \delta Z) + \mathcal{O}(\epsilon^2)~.
\end{split}
\end{equation}
Up to $\mathcal{O}(\epsilon)$, we can then express $\delta Z$ as
\begin{equation}
    \delta Z = \sum_{n=1}^{N-2} \frac{(-1)^n}{r^n} \sum_{m=0}^n H_{n,m} B_{(m)}~.
\end{equation}
In the subsequent calculations, we have omitted the $\ell$ from the subscripts of $\chi$ or $\Phi$ for brevity of notation. Following the algorithm in \cite{Katagiri:2023umb} to cast the wave equation into a parametrized form, one can redefine the field $\chi(r)$ as
\begin{equation}
    \chi(r) \to \Phi(r) \equiv \chi ~ \sqrt{1+\delta Z}~.
\end{equation}

The resultant wave equation for $\Phi(r)$ can then be written in the parametrized form given by
\begin{equation}\label{eq:parwaveeqstat}
    \begin{split}
        & \bar f~ \partial_r (\bar f~ \partial_r \Phi) - \bar f~ V~\Phi = 0 \\
        &V = V_{BG} + \delta \tilde V 
        ~~,~~ V_{BG} = \frac{\ell (\ell+1)}{r^2} + \frac{1}{r} \partial_r \bar f 
        ~~,~~ \delta \tilde V = \frac{1}{r_+^2} \sum_{j=3}^{N+2} \alpha_j \left(\frac{r_+}{r}\right)^j~.
    \end{split}
\end{equation}
The effective potential $V(r)$ has been split into the background contribution: $V_{BG}(r)$, which contains the contribution of the Ladder-symmetric background spacetime with modified horizon radii $r_+$ and $r_-$; and, $\delta \tilde V(r)$ which contains the information of the deviations away from this background in the parametrized form. The parameters $\alpha_j$ can be derived to be
    \begin{equation}
    \begin{split}
        \alpha_j =& \sum_{m=0}^{j-2} \frac{(-1)^j}{2r_+^{j-2}} H_{j-2,m} B_{(m)} \left\{  (j-2)(j-3) - 2\ell(\ell+1)\right\} \\
        &- \sum_{m=0}^{j-3} \frac{(-1)^j}{2r_+^{j-2}} H_{j-3,m} (r_+ + r_-) (j-3)^2 B_{(m)}  + \sum_{m=0}^{j-4} \frac{(-1)^j}{2r_+^{j-2}} H_{j-4,m} (j-3)(j-4) (r_+ r_-) B_{(m)}~.
    \end{split}
    \end{equation}
\end{widetext}
The solution of the field $\Phi(r)$ can be cast into a similar form as
\begin{equation}
    \Phi(r) = \Phi^{(0)}(r) + \sum_{j=3}^{N+2} \alpha_j \Phi^{(1)}_j (r)~.
\end{equation}
\begin{widetext}
At $\mathcal{O}(\alpha_j^0)$, the equation for $\Phi^{(0)}(r)$ is given by
\begin{equation}\label{eq:wavesphericalbg}
    r^2 \left(1 - \frac{r_+}{r}\right)\left(1 - \frac{r_-}{r}\right) \partial_r^2 \Phi^{(0)}(r) + r \left( \frac{r_+ + r_-}{r} - \frac{2 r_+ r_-}{r^2}  \right)   \partial_r \Phi^{(0)}(r)  - \left( \ell(\ell+1) + \frac{r_+ + r_-}{r} - \frac{2 r_+ r_-}{r^2}\right) \Phi^{(0)}(r)  = 0~.
\end{equation}
From this, $\Phi^{(0)}(r)$ can be solved directly in terms of Legendre polynomials. Using the boundary conditions of regularity at the horizon and that at $r >> r_+, \Phi^{(0)}(r) = (r/r_+)^{\ell+1}+ O (r/r_+)^{\ell}$ 

\begin{equation}
    \Phi^{(0)}(r) = \frac{(\ell!)^2}{(2\ell)!}~ \left(1-\frac{r_-}{r_+}\right)^\ell ~ \left(\frac{r}{r_+}\right)~ P_\ell \left(\frac{(r-r_+) + (r-r_-)}{(r_+-r_-)}\right)~,
\end{equation}
\end{widetext}
where $P_\ell\left(\frac{(r-r_+) + (r-r_-)}{(r_+-r_-)}\right)$ is the Legendre polynomial of the first kind of order $\ell$. For convenience in further calculations, we will express $\Phi^{(0)}(r)$ as a finite series in $(r_+/r)$ given by
\begin{equation}
    \Phi^{(0)}(r) = \left( \frac{r}{r_+} \right)^{\ell+1} \sum_{n=0}^\ell \phi_0^{(n)} \left( \frac{r_+}{r} \right)^{n}~.
\end{equation}
From the form of $\Phi^{(0)}(r)$, it is evident that there is no contribution at $\mathcal{O}(r_+/r)^\ell$. Hence, for the background solution, TLN vanishes. This is expected as we are working with a Ladder-symmetric background.
\begin{widetext}
The equation for $\Phi^{(1)}_j$ is given by the $\mathcal{O}(\alpha_j^1)$ part of the wave equation as
\begin{equation}\label{eq:wavesphericalalpha}    (r-r_+)(r-r_-)~\partial_r^2 \Phi^{(1)}_j + \left[ (r_+ + r_-) - \frac{2r_+ r_-}{r} \right] \partial_r \Phi^{(1)}_j - \left[ \ell(\ell+1) + \frac{r_+ + r_-}{r} - \frac{2r_+ r_-}{r^2} \right] \Phi^{(1)}_j = \left( \frac{r_+}{r} \right)^{j-3-\ell} \sum_{n=0}^\ell \phi_0^{(n)} \left( \frac{r_+}{r} \right)^{n}~.
\end{equation}
We express $\Phi^{(1)}_j$ as an infinite series in $(r_+/r)$, in the same manner as we did for $\Phi^{(0)}$,
\begin{equation}
    \Phi^{(1)}_j = \left( \frac{r}{r_+} \right)^{\ell+1} \sum_{n=0}^\infty \gamma_n \left( \frac{r_+}{r} \right)^{n} = \sum_{n=0}^\infty \gamma_n \left( \frac{r_+}{r} \right)^{n-\ell-1}~.
\end{equation}
Substituting this expansion in eq. (\ref{eq:wavesphericalalpha}) and dividing by a factor of $(r_+/r)^\ell$ on both sides for simplicity of expression, we get a recurrence relation for the $\gamma_n$'s.
    \begin{equation}\label{eq:recrelfull}
    \begin{split}
        &\sum_{n=0}^\infty \gamma_n \left[ \left( \frac{r_+}{r} \right)^{n-1} \left\{ n(n-2\ell-1) \right\} - \left( \frac{r_+}{r} \right)^{n} (n-\ell)^2  \left(1 + \frac{r_-}{r_+} \right) +  \left( \frac{r_+}{r} \right)^{n+1} \left\{ (\ell-n) (\ell-n-1) \right\}  \left( \frac{r_-}{r_+} \right) \right] \\
        &= \sum_{n=0}^\ell  \left( \frac{r_+}{r} \right)^{j-3+n} \phi_0^{(n)}~.
    \end{split}
    \end{equation}
\end{widetext}
From this recurrence relation, two possible cases can arise.
\begin{enumerate}
    \item \textbf{Case 1: }$3 \leq j \leq 2\ell+3$. In this case, there are contributions from logarithmic terms in the TLNs, and eq. (\ref{eq:recrelfull}) needs to be solved directly to obtain the TLNs.
    \item \textbf{Case 2: } $2\ell+4 \leq j$. In this case, it is possible to solve for the $\gamma_n$'s in eq. (\ref{eq:recrelfull}) by comparing the coefficients on both sides of the equation.
\end{enumerate}
Due to the computational complexities involved, it is difficult to derive generic $(\ell,j)$ dependent forms of the TLNs. In appendix \ref{app:basisTLNsnonrunning}, we outline the solutions to the $\gamma_n$'s in eq. (\ref{eq:recrelfull}) in terms of $\gamma_{2\ell+1}$.

However, note that the expression for the TLN at any value of $\ell$ or $N$ gives us a linear equation on $\beta_{(n)}$'s. Now, this finite number of $\beta_{(n)}$'s depend only on the properties of the underlying spacetime, and are independent of the multipole moment of the perturbation being considered. But, if we demand that the TLNs vanish identically for every $\ell$, the constraint equations on these parameters are infinite in number, and consist of coefficients which are explicitly dependent on $\ell$. Therefore, a general solution of the $\beta_{(n)}$'s must depend on $\ell$'s if all of the infinite number of constraint equations are to be solved. This contradicts our assumption that the $\beta_{(n)}$'s are independent of $\ell$. Therefore, the only consistent solution to this would be to set $\beta_{(n)}  = 0 ~\forall~ n \in ~ [3, N]$, which gets us back to the Ladder-symmetric background metric. Hence, any deviation from the Ladder-symmetric background metric would result in at least some TLN being non-zero, and this proves our statement that the Ladder-symmetry is a necessary condition for TLNs to vanish for all values of $\ell$.

As an illustration of this argument, we have derived specific expressions of the bases and TLNs for two different cases in eq. (\ref{eq:statbggeneral}), $N=4$ in Sec. \ref{subsec:n=4} and $N=6$ in Sec. \ref{subsec:n=6}, and show that demanding the running contribution to the TLNs to vanish for all values of $\ell$ eliminates all deviations from the Ladder-symmetric background metric. This can, in principle, be generalized to all other $N$ values.

\subsection{$N=4$ in $f(r)$}\label{subsec:n=4}

In this case, we have added two additional parameters into the metric, viz. $\beta_{(3)},$ and $\beta_{(4)}$. 

To check whether there can exist some non-zero value of $\beta_{(3)}$ and $\beta_{(4)}$ for which the running contribution to the TLNs can still be zero, we need to supply at least two equations. Here, we will use the equations of the running contributions to the TLNs at $\ell=1$ and $\ell=2$. For $N=4$, we get non-zero $\alpha_j$'s for $j \in [3,6]$. At $\ell =1,$ the TLN gets a log-running contribution from $\alpha_3, \alpha_4,$ and $\alpha_5$, and for $\ell=2$, all of the $\alpha_j$'s contribute to the running part of the TLN only. The $\alpha_j$'s have the following values:
\begin{equation}
    \begin{split}
        \alpha_3 &= -\epsilon\frac{\beta_{(3)}\ell(\ell+1) }{r_+^2 r_-}-\epsilon\frac{\beta_{(4)}\ell(\ell+1)  (r_++r_-)}{r_+^3 r_-^2} \\
        \alpha_4 &=-\epsilon\frac{\beta_{(3)}   (r_++r_-)}{2 r_+^3 r_-}  -\epsilon\frac{\beta_{(4)}   \left(2\ell(\ell+1)r_+ r_-+r_+^2+r_-^2\right)}{2 r_+^4 r_-^2}\\
        \alpha_5 &=\epsilon \frac{\beta_{(3)}  }{r_+^3}- \epsilon \frac{\beta_{(4)}   (r_++r_-)}{r_+^4 r_-} ~,~~
        \alpha_6 = \epsilon \frac{3 \beta_{(4)}  }{r_+^4}~.\\
    \end{split}
\end{equation}
The bases can be calculated for $\ell = 1$ to be
\begin{equation}\label{eq:l1basis}
    \begin{split}
        d_3 &= -\frac{(r_++r_-)^2}{12 r_+^2} ~,~~
        d_4 = \frac{r_++r_-}{3 r_+} ,~~
        d_5 = - \frac{1}{3}~,
    \end{split}
\end{equation}
and for $\ell=2$ we get,
\begin{equation}\label{eq:l2basis}
    \begin{split}
        d_3 &= -\frac{\left(r_+^2+4 r_+ r_-+r_-^2\right)^2}{180 r_+^4} \\
        d_4 &= \frac{(r_++r_-) \left(r_+^2+4 r_+ r_-+r_-^2\right)}{15 r_+^3} \\
        d_5 &= -\frac{2 (2 r_++r_-) (r_++2 r_-)}{15 r_+^2} ~,~
        d_6 = \frac{2 (r_++r_-)}{5 r_+}~.
    \end{split}
\end{equation}
Using these, the running contributions to the TLN can be calculated to be
\begin{equation}
    \begin{split}
        \text{For $\ell=1$: } &\sum_{j=3}^5 \alpha_j d_j =  -\epsilon\frac{\beta_{(3)}  }{3 r_+^3} \\
        \text{For $\ell=2$: } &\sum_{j=3}^6 \alpha_j d_j = -\frac{\epsilon  (r_++r_-) \left(\beta_{(3)} (r_++r_-)-4 \beta_{(4)}\right)}{5 r_+^5}~.
    \end{split}
\end{equation}
Setting these two expressions to zero, the only possible consistent solution turns out to be
\begin{equation}
\beta_{(3)} = \beta_{(4)} = 0~.
\end{equation}
Consequently, this makes all the $\alpha_j$'s vanish. Then, due to $\alpha_6$ becoming zero, even the non-running contribution to the TLN at $\ell=1$ vanishes.

From this, we infer that if $f(r)$ is modified from the Ladder-symmetric background value in a series of inverse powers of $r$ at most up to $r^{-4}$, then it will necessarily lead to non-zero TLNs, which can be observed in $\ell$-values as low as $1$ or $2$. Conversely, if we allow perturbative modifications in $f(r)$ at most up to $r^{-4}$, then the only possibility to have vanishing TLNs for all values of $\ell$ is to have a Ladder-symmetric background with $f(r) = (1-R_+/r)(1-R_-/r)$.

\subsection{$N=6$ in $f(r)$}\label{subsec:n=6}

A similar check can be done if we add now four additional parameters $\left(\beta_{(3)}, \beta_{(4)}, \beta_{(5)} \,\text{and}\, \beta_{(6)}\right)$ to the metric. To fix these four parameters, we will need at least four equations, that is, the running contribution equations from $\ell = 1, 2, 3 \,\text{and}\,4$. We get non-zero $\alpha_j$'s for $j \in [3,8]$ for $N=6$. Only the first three $\alpha_j$'s will contribute to the running part of the TLNs at $\ell = 1$, the first five $\alpha_j$'s will contribute at $\ell = 2$, and all of them will contribute to the running part for $\ell = 3$ and $\ell = 4$. The $\alpha_j$'s for $N=6$ will have the following expression:
\begin{widetext}
\begin{equation}
    \begin{split}
        \alpha_3 =& -\epsilon\frac{\beta_{(3)}\ell(\ell+1) }{r_+^2 r_-}-\epsilon\frac{\beta_{(4)}\ell(\ell+1)  (r_++r_-)}{r_+^3 r_-^2} - \epsilon \frac{\beta_{(5)}\ell(\ell+1)(r_+^2 + r_+ r_- + r_-^2)}{r_+^4 r_-^3}- \epsilon \frac{\beta_{(6)}\ell(\ell+1)(r_+ + r_-)(r_+^2 + r_-^2)}{r_+^4 r_-^3}\\
        \alpha_4 =&-\epsilon\frac{\beta_{(3)}   (r_++r_-)}{2 r_+^3 r_-}  -\epsilon\frac{\beta_{(4)}   \left(2\ell(\ell+1)r_+ r_-+r_+^2+r_-^2\right)}{2 r_+^4 r_-^2}-\epsilon \frac{\beta_{(5)}(r_++r_-)\left((2\ell(\ell+1)-1)r_+ r_-+r_+^2+r_-^2\right)}{2 r_+^5 r_-^3} \\
        &- \epsilon \frac{\beta_{(6)}\left(2\ell(\ell+1)r_+ r_-\left(r_+^2 + r_+ r_- + r_-^2\right)+r_+^4+r_-^4\right)}{2 r_+^6 r_-^4}\\
        \alpha_5 =&\,\epsilon \frac{\beta_{(3)}  }{r_+^3}- \epsilon \frac{\beta_{(4)}   (r_++r_-)}{r_+^4 r_-} -\epsilon \frac{{\beta}_{(5)} \left(\ell(\ell+1) r_+ r_-+r_+^2+r_-^2\right)}{r_+^5 r_-^2}-\epsilon \frac{\beta_{(6)} \left(\ell(\ell+1) \left(r_+^2 r_- + r_+ r_-^2\right)+r_+^3+r_-^3\right)}{r_+^6 r_-^3}\\
        \alpha_6 =&\, \epsilon \frac{3 \beta_{(4)}}{r_+^4}-\epsilon \frac{3 \beta_{(5)} (r_++r_-)}{2 r_+^5 r_-} -\epsilon\frac{\beta_{(6)}  \left(2 \ell(\ell+1) r_+ r_-+3 r_+^2+3 r_-^2\right)}{2 r_+^6 r_-^2}~,~
        \alpha_7 =\, \epsilon \frac{6 \beta_{(5)} }{r_+^5}-\epsilon  \frac{2 \beta_{(6)}(r_++r_-)}{r_+^6 r_-}~,~
        \alpha_8 =\, \epsilon \frac{10 \beta_{(6)}}{r_+^6}~.
    \end{split}
\end{equation}

The bases for $\ell=1$ and $\ell=2$ would be the same as in eq. (\ref{eq:l1basis}) and eq. (\ref{eq:l2basis}) respectively. For $\ell = 3$, we can calculate the bases as

\begin{equation}\label{eq:l3basis}
    \begin{split}
        d_3 &= -\frac{(r_++r_-)^2 \left(r_+^2+8 r_+ r_-+r_-^2\right)^2}{2800 r_+^6} \\
        d_4 &= \frac{3 (r_++r_-) \left(r_+^2+3 r_+ r_-+r_-^2\right) \left(r_+^2+8 r_+ r_-+r_-^2\right)}{350 r_+^5} \\
        d_5 &= -\frac{51 r_+^4+366 r_+^3 r_-+666 r_+^2 r_-^2+366 r_+ r_-^3+51 r_-^4}{700 r_+^4} \\
        d_6 &= \frac{(r_++r_-) \left(19 r_+^2+62 r_+ r_-+19 r_-^2\right)}{70 r_+^3} \\
        d_7 &= -\frac{3 \left(23 r_+^2+54 r_+ r_-+23 r_-^2\right)}{140 r_+^2} \\
        d_8 &= \frac{3 (r_++r_-)}{7 r_+}~,
    \end{split}
\end{equation}

and for $\ell=4$ as : 
\begin{equation}\label{eq:l4basis}
    \begin{split}
        d_3 &= -\frac{\left(r_+^4+16 r_+^3 r_-+36 r_+^2 r_-^2+16 r_+ r_-^3+r_-^4\right)^2}{44100 r_+^8} \\
        d_4 &= \frac{2 (r_++r_-) \left(r_+^2+5 r_+ r_-+r_-^2\right) \left(r_+^4+16 r_+^3 r_-+36 r_+^2 r_-^2+16 r_+ r_-^3+r_-^4\right)}{2205 r_+^7} \\
        d_5 &= -\frac{29 r_+^6+408 r_+^5 r_-+1677 r_+^4 r_-^2+2632 r_+^3 r_-^3+1677 r_+^2 r_-^4+408 r_+ r_-^5+29 r_-^6}{2205 r_+^6} \\
        d_6 &= \frac{2 (r_++r_-) \left(97 r_+^4+802 r_+^3 r_-+1632 r_+^2 r_-^2+802 r_+ r_-^3+97 r_-^4\right)}{2205 r_+^5} \\
        d_7 &= -\frac{692 r_+^4+4232 r_+^3 r_-+7302 r_+^2 r_-^2+4232 r_+ r_-^3+692 r_-^4}{2205 r_+^4} \\
        d_8 &= \frac{4 (r_++r_-) (5 r_++2 r_-) (2 r_++5 r_-)}{63 r_+^3}~.
    \end{split}
    \end{equation}
Using these, the running contributions of the TLNs can be written as
\begin{equation}
    \begin{split}
        \text{For $\ell=1$: } &\sum_{j=3}^5 \alpha_j d_j =  -\epsilon\frac{\beta_{(3)}  }{3 r_+^3} \\
        \text{For $\ell=2$: } &\sum_{j=3}^7 \alpha_j d_j = -\epsilon \frac{\left(\beta_{(3)} (r_++r_-)^2-4 \beta_{(4)} (r_++r_-)+4 \beta_{(5)}\right)}{5 r_+^5}\\
        \text{For $\ell=3$: } &\sum_{j=3}^8 \alpha_j d_j = \epsilon \frac{9}{175 r_+^7} \big[-\beta_{(3)} \left(r_+^2+3 r_+ r_-+r_-^2\right)^2 + 10 \beta_{(4)} (r_++r_-) \left(r_+^2+3 r_+ r_-+r_-^2\right) \\
        & ~~~~~~~~~~~~~~~~~~~~~~~~~~\,-5 \beta_{(5)} \left(7 r_+^2+16 r_+ r_-+7 r_-^2\right)+50 \beta_{(6)} (r_++r_-) \big]\\
        \text{For $\ell=4$: } &\sum_{j=3}^8 \alpha_j d_j = \epsilon  \frac{4}{441 r_+^9} \big[- \beta_{(3)} (r_++r_-)^2 \left(r_+^2+5 r_+ r_-+r_-^2\right)^2 \\
        &~~~~~~~~~~~~~~~~~~~~~~~~~~\,+6 \beta_{(4)}(r_++r_-) \left(r_+^2+5 r_+ r_-+r_-^2\right) \left(3 r_+^2+8 r_+ r_-+3 r_-^2\right) \\
        &~~~~~~~~~~~~~~~~~~~~~~~~~~\,-3 \beta_{(5)} \left(41 r_+^4+242 r_+^3 r_-+414 r_+^2 r_-^2+242 r_+ r_-^3+41 r_-^4\right) \\
        &~~~~~~~~~~~~~~~~~~~~~~~~~~\,+14 \beta_{(6)} (r_++r_-) \left(29 r_+^2+82 r_+ r_-+29 r_-^2\right)\big]~.
    \end{split}
\end{equation}
\end{widetext}

Again, the only consistent solution that sets all the above expressions to zero is
\begin{equation}
\beta_{(3)} = \beta_{(4)} = \beta_{(5)} = \beta_{(6)} = 0~,
\end{equation}
which makes all the $\alpha_j$'s vanish and returns us the Ladder-symmetric metric.

\section{Parametrized Love numbers   for Stationary, Rotating Metrics of the KRZ Class}\label{sec:TLNrotating}

In this section, we derive the parametrized TLNs for the general rotating KRZ class of solutions. We follow the same steps as those in the last section, and show that to have vanishing TLNs for all $\ell \geq 1$, all the deviations from the Ladder-symmetric background metric must reduce to zero. As discussed in Sec. \ref{subsec:strategy}, we can set $R_B=1$ in eq. (\ref{eq:genKRZmet}) by utilizing the residual gauge freedom, and the resultant wave equation becomes
\begin{equation}
    \begin{split}
        -\Delta(r) \Bigg[ \partial_r \Big\{\Delta(r) \partial_r \Big\}  - \ell(\ell+1) + \frac{a^2 m^2}{\Delta(r)} \Bigg] \chi(r) &= 0~,
    \end{split}
\end{equation}
with
\begin{equation}
    \Delta(r)=r^2 f_0(r) = r^2 R_\Sigma - r~R_M + a^2~.
\end{equation}
In Appendix \ref{app:KRZTLN0}, we show that the TLNs for a black hole belonging to this class of spacetimes are zero for all values of $\ell$ and $m$.

On the Ladder-symmetric background, we now add perturbations to $f(r)$ of the form
\begin{equation}
    f(r) = f_0(r) +\epsilon \sum_{n=3}^N \frac{\beta_n}{r^n}~,
\end{equation}
where $f_0(r)$ is the same as in eq. (\ref{eq:genf0}). The $\beta_n$'s in $f(r)$ can arise due to sub-leading terms in $(1/r)$ in both $R_\Sigma(r)$ and $R_M(r)$. For $m=0$, the last term in the effective potential vanishes, resulting in an identical form of the wave equation for both the rotating $a\neq 0$ case and the static spherically symmetric case discussed in Sec. \ref{sec:TLNsphsym}. Hence, the same analysis as in the previous section follows through in the stationary rotating KRZ class of spacetimes, and we can infer the necessity of Ladder-symmetry for the vanishing of static scalar TLNs in this case as well. 

For completeness, we will calculate the parameters $\alpha_j$'s in this section for the stationary KRZ class of spacetimes rotating with $a = (J/M^2) \neq 0$, and as a consistency check, we will show that their form matches those derived in Sec. \ref{sec:TLNsphsym} at the limits $a \to 0$ or $m \to 0$.

Modifying the effective potential from $f_0(r)$ to $f(r)$ leads to an infinite series due to the presence of $(r^4 f)^{-1}$ like terms. Writing 
\begin{equation}
    f(r) = \bar f(r) \prod_{n=3}^N \left( 1- \frac{r_n}{r} \right) = \bar f(r) (1+\delta Z)~,
\end{equation}
as in the previous section, we can express $\delta Z$ as 
\begin{equation}
    \delta Z = \sum_{n=1}^{N-2} \frac{(-1)^n}{r^n} \sum_{m=0}^n H_{n,m} B_{(m)}~,
\end{equation}
and therefore express $(r^4 f)^{-1}$ as
\begin{equation}
    \begin{split}
        \frac{1}{r^4 f} &= \frac{1}{r^4 \bar f} (1 + \delta Z)^{-1}  \equiv  \frac{1}{r^4 \bar f}\left( 1 + \sum_{n=1}^\infty \frac{\xi_n}{r^n}\right)~.
    \end{split}
\end{equation}
Here, the $\xi_n$'s are symbolic representations of the coefficients of the infinite power series $(1+\delta Z)^{-1}$, that can be read off from the expansion. One can further calculate $(r^4 \bar f^{-1})$ as
\begin{equation}
    \begin{split}
        &\frac{1}{r^4 \bar f} (1 + \delta Z)^{-1} \\
        &= \frac{1}{r^4 \left(1 - \frac{r_+}{r}\right)\left(1 - \frac{r_-}{r}\right)} \left[ 1 + \sum_{n=1}^\infty \frac{\xi_n}{r^n} \right] \\
        &= \frac{1}{r^4 \bar f} + \frac{1}{r^4} \left[ \sum_{n_1 =0}^\infty \left( \frac{r_+}{r}\right)^{n_1} \right] \left[ \sum_{n_2 =0}^\infty \left( \frac{r_-}{r}\right)^{n_2} \right] \left[ \sum_{n_3 =1}^\infty \frac{\xi_{n_3}}{r^{n_3}} \right] \\
        &\equiv \frac{1}{r^4 \bar f} + \frac{1}{r^4} \sum_{n=1}^\infty \frac{\rho_n}{r^n},\text{ where we define }\\
        &\sum_{n=1}^\infty \frac{\rho_n}{r^n} \equiv \left[ \sum_{n_1 =0}^\infty \left( \frac{r_+}{r}\right)^{n_1} \right] \left[ \sum_{n_2 =0}^\infty \left( \frac{r_-}{r}\right)^{n_2} \right] \left[ \sum_{n_3 =1}^\infty \frac{\xi_{n_3}}{r^{n_3}} \right]   \\
        &\Rightarrow \left( \frac{a^2 m^2}{r^4 f} - \frac{a^2 m^2}{r^4 \bar f} \right) = \sum_{n=1}^\infty \frac{a^2 m^2 \rho_n}{r^{n+4}}
    \end{split}
\end{equation}
Again, $\rho_n$ is a symbolic representation like $\xi_n$ that can be read off from the infinite series expansions. It is important to note here that, since all the coefficients $\rho_n$'s or $\xi_n$'s depend at least linearly on the sums of the products of the perturbative roots, they only exist when at least one of the $\beta_n$s in $f(r)$ is non-zero. 

Representing another infinite series
\begin{equation}
    \frac{1}{\bar f}~\delta Z \equiv \sum_{n=1}^\infty \frac{\lambda_n}{r^n}
\end{equation}
in a similar way, we can finally express the wave equation for the field $\chi = (\Phi/r)$ on the perturbed spacetime as
\begin{equation}
    \begin{split}
        &\bar f ~ \partial_r ( \bar f ~\partial_r \Phi ) - \bar f~ \left(V_{BG} + \delta \tilde V\right)~ \Phi =0 \\
        &\text{where } V_{BG} = \frac{\ell(\ell+1)}{r^2} + \frac{1}{r} \partial_r \bar f - \frac{a^2 m^2}{r^4 \bar f} \\
        &\text{and }\delta \tilde V = \frac{1}{r_+^2} \sum_{j=3}^\infty \alpha_j \left(\frac{r_+}{r}\right)^j~,
    \end{split}
\end{equation}
with $\alpha_j$ given by
\begin{equation}
\begin{split}
    \alpha_j = &\sum_{m=0}^{j-2} \frac{(-1)^j}{2r_+^{j-2}} H_{j-2,m} B_{(m)} \left\{  (j-2)(j-3) - 2\ell(\ell+1)\right\} \\
        &- \sum_{m=0}^{j-3} \frac{(-1)^j}{2r_+^{j-2}} H_{j-3,m} (r_+ + r_-) (j-3)^2 B_{(m)}  \\
        &+ \sum_{m=0}^{j-4} \frac{(-1)^j}{2r_+^{j-2}} H_{j-4,m} (j-3)(j-4) (r_+ r_-) B_{(m)} \\
        &- \frac{a^2 m^2}{(r_+)^{j-4}} (\rho_{j-4} + \lambda_{j-4})~.
\end{split}
\end{equation}
It is interesting to notice here that, schematically,
\begin{equation}\label{eq:alpharot}
    (\alpha_j)_{\text{rotating}} \sim (\alpha_j)_{\text{static}} - \frac{a^2 m^2}{(r_+)^{n-2}}  (\rho_{j-4} + \lambda_{j-4})~.
\end{equation}
Therefore, for $m=0$, we find that \begin{equation}
    (\alpha_j)_{\text{rotating}}  \sim (\alpha_j)_{\text{static}}~,
\end{equation}
and hence, the infinite series in $(r_+/r)$ in the effective potential truncates to a finite number.

As mentioned before, for axisymmetric perturbations ($m=0$), the perturbative corrections for the effective potential have the same form as in the static spherically symmetric spacetime discussed in Sec. \ref{sec:TLNsphsym}. Hence, the same analysis follows through in this case, and we conclude that the existence of Ladder symmetry is necessary for the vanishing of TLNs for all values of $\ell$ in this class of spacetimes as well.

\section{Conclusions and Discussions}\label{sec:conclude}

In this study, we have demonstrated that the presence of Ladder symmetry is both a necessary and sufficient condition for the vanishing of static tidal Love numbers in black holes, encompassing both static and rotating configurations within the KRZ parametrization. Any perturbative breaking of this symmetry inevitably induces nonzero scalar TLNs, signifying that the “no-Love” property is lost. Imposing the requirement that the TLNs vanish for all multipole moments $\ell$ yields an infinite hierarchy of interlinked constraints among the deformation parameters. The only self-consistent resolution of this system is achieved when all deformation parameters identically vanish, thereby reinstating the exact Ladder symmetry. Hence, Ladder symmetry stands as the defining and indispensable criterion underlying the “no-Love” condition for scalar perturbations.

The result extends naturally beyond the KRZ class. In more general or non-separable spacetimes, one may expect generalized Ladder-like structures that approximately suppress, but do not eliminate, tidal responses.
Recent works like \cite{Berens:2025jfs,Berens:2025okm} study the vanishing of tidal response for higher-spin perturbations in different dimensions using similar symmetries. It would be interesting to investigate whether, in those setups as well, the existence of these symmetries is a necessary criterion for the vanishing of TLNs. The vanishing of tidal response at a full nonlinear order has also been reported in \cite{Parra-Martinez:2025bcu} for spherically symmetric black holes in asymptotically flat general relativity using worldline EFT. It would be interesting to use such worldline EFT calculations in our framework to understand the existence of running static scalar TLNs on spacetimes perturbatively deviated from Ladder-symmetric KRZ. In \cite{Chakraborty:2025zyb}, the authors have shown that the TLNs of a black hole perturbed by fermionic fields are zero. It would be interesting to probe it from a symmetry perspective and check whether it is possible to show that there never exists a ``Ladder-like" symmetry in such cases, which would then correspond to non-vanishing TLNs for all perturbation modes. It would also be of interest to move beyond the KRZ parametrization to explore the manifestation of Love symmetry in more general stationary black hole spacetimes. Such an endeavor may also demand a deeper generalization of the very structure of the Love symmetry itself. 

Interestingly, our analysis suggests that if one were to include an infinite hierarchy of correction terms in the metric, the tidal Love numbers (TLNs) could, in principle, vanish only through a highly fine-tuned cancellation among all coefficients. Any finite truncation of this expansion, however, inevitably fails to satisfy the complete set of constraints, leaving residual, nonzero TLNs. Even in the limit of infinitely many corrections, achieving a strictly vanishing TLN would require the deformation parameters to depend intricately on the specific black hole parameters, rather than being universal features common to all black holes. From an astrophysical perspective, such delicate cancellations appear implausible: once the underlying symmetry is even slightly broken, TLNs are expected to emerge for all multipole orders $\ell$. Consequently, the detection of even a small but nonzero TLN would signal a genuine breaking of the Ladder symmetry and thus indicate a true deviation from this class of black hole geometries.

\section*{Acknowledgement}

We would like to express our heartfelt gratitude to Soham Acharya, Rajes Ghosh, Sreejith Nair, and Sumanta Chakraborty for their valuable discussions and helpful inputs. We also thank Lam Hui, Luca Santoni, and Alessandro Podo for their useful comments on the results during the ``Theoretical Tools for Gravitational Waves Physics" workshop organized in ETH Zurich, where the results were presented by SR. SS’s research is supported by the Department of Science and Technology, Government of India, under the ANRF CRG Grant (No. CRG/2023/000934).

\appendix

\section*{Appendix}

\section{Form of $f_0(r)$ for the restricted class of KRZ metrics from \cite{Sharma:2024hlz}}\label{app:LaddSymKRZ}

In this appendix, we'll identify the specific terms of $R_\Sigma(r)$ and $R_M(r)$ that contribute to the expression of $f_0(r)$. Suppose, the functions $R_\Sigma$ and $R_M$ can be written as infinite summations
\begin{equation}
    R_{\Sigma,M} = \sum_{-\infty}^\infty R_{\Sigma,M}^{(n)} ~r^n~.
\end{equation}
In general, the functions $R_{\Sigma, M}$ can have any arbitrary form. But for our purpose here, we know that the $f_0(r)$ we would be working with allows only a power series of $\frac{1}{r}$. Hence, although such an infinite series form of $R_{\Sigma, M}$ might not be applicable for some forms (e.g., those including logarithmic terms), it suffices for our purposes in this work.

Following \cite{Sharma:2024hlz}, $f_0$ can be written as
\begin{equation}
    f_0(r) = R_\Sigma - \frac{R_M}{r} + \frac{a^2}{r^2}~.
\end{equation}
Imposing Ladder symmetry, one finds that the form of $f_0(r)$ is constrained to be
\begin{equation}
     f_0(r) = 1 + \frac{A}{r} + \frac{B}{r^2} = \left( 1- \frac{R_+}{r} \right) \left( 1- \frac{R_-}{r} \right)~.
\end{equation}
Comparing the two expressions of $f_0$, one finds that the $R_{\Sigma,M}^{(n)}$s are constrained as
\begin{align}
        &R_\Sigma^{(0)} - R_M^{(1)} = 1,\\
        &R_\Sigma^{(-1)} - R_M^{(0)} = A, \text{ and}\\
        &R_\Sigma^{(-2)} - R_M^{(-1)} + a^2 = B; \\
        &R_\Sigma^{(n)} = 0 \text{ for $n<-2$ and $n>0$},\\
        &R_M^{(n)} = 0 \text{ for $n<-1$ and $n>1$}~.
\end{align}
Hence, the form of $f_0$ can be expressed as
\begin{equation}
    f_0(r) = 1 + \frac{R_\Sigma^{(-1)} - R_M^{(0)}}{r} + \frac{R_\Sigma^{(-2)} - R_M^{(-1)} + a^2}{r^2}~.
\end{equation}
Up to this point, we have not yet assumed the existence of real roots of $f_0(r)$, or any criteria about the existence or any properties of the horizon. Such assumptions would lay further restrictions on the numbers $R_{\Sigma, M}^{(n)}$, but we would not require these details for casting the effective potential into the parametrized form.

\section{Product of Sums to Sum of Products: A closed form}\label{app:POStoSOP}

In this appendix, we'll demonstrate how our method of converting the infinite product into an infinite summation helps us cast the effective potential into a parametrized form as in \cite{Cardoso:2019mqo}, and how it should help us in detecting deviations from Ladder symmetry.

The product
\begin{equation}
    \prod_{n=3}^N  \left(1 - \frac{r_n}{r}\right) = \left(1 - \frac{r_3}{r}\right)~ \left(1 - \frac{r_4}{r}\right)~ \cdots ~\left(1 - \frac{r_N}{r}\right)
\end{equation}
has $(N-2)$ factors, and can be written as a power series in $\frac{1}{r}$ as
\begin{equation}
    1 - \frac{1}{r} \sum_{n=3}^N r_n + \frac{1}{r^2} \sum_{n_1=3}^{N-1} \sum_{n_2=n_1+1}^N r_{n_1} r_{n_2} + \cdots + (-1)^{N-2} \frac{1}{r^{N-2}} \prod_{n=3}^{N} r_n~.
\end{equation}
This massive summation can be expressed compactly as a nested summation of products of $r_n$s as
\begin{equation}
    \sum_{k=0}^{N-2} \frac{(-1)^k}{r^k} \sum_{i_1 = 3}^{N-(k-1)} \sum_{i_2 = i_1+1}^{N-(k-2)} \cdots \sum_{i_k = i_{k-1} + 1}^{N} \prod_{j=1}^k r_{i_j}~.
\end{equation}
This is the full expression of $(1+\delta Z)$. Next, we can separate out the $k=0$ term of this summation and extract $\delta Z$ from this as
\begin{widetext}
   \begin{equation}
    \begin{split}
        \prod_{n=3}^N  \left(1 - \frac{r_n}{r}\right) &= 1 + \sum_{k=1}^{N-2} \frac{(-1)^k}{r^k} \sum_{i_1 = 3}^{N-(k-1)} \sum_{i_2 = i_1+1}^{N-(k-2)} \cdots \sum_{i_k = i_{k-1} + 1}^{N} \prod_{j=1}^k r_{i_j} = 1 + \delta Z \\
        \Rightarrow \delta Z &= \sum_{k=1}^{N-2} \frac{(-1)^k}{r^k} \sum_{i_1 = 3}^{N-(k-1)} \sum_{i_2 = i_1+1}^{N-(k-2)} \cdots \sum_{i_k = i_{k-1} + 1}^{N} \prod_{j=1}^k r_{i_j} \\
        &\equiv \sum_{k=1}^{N-2} \frac{(-1)^k}{r^k} A_k ~,~~
        \text{with }A_k \equiv \sum_{i_1 = 3}^{N-(k-1)} \sum_{i_2 = i_1+1}^{N-(k-2)} \cdots \sum_{i_k = i_{k-1} + 1}^{N} \prod_{j=1}^k r_{i_j}~.
    \end{split}
\end{equation}

On taking the limit $N \to \infty$, the product becomes
\begin{equation}
    \begin{split}
        \prod_{n=3}^\infty  \left(1 - \frac{r_n}{r}\right) &= 1 + \sum_{k=1}^\infty \frac{(-1)^k}{r^k} \sum_{i_1 = 3}^\infty \sum_{i_2 = i_1+1}^\infty \cdots \sum_{i_k = i_{k-1} + 1}^\infty \prod_{j=1}^k r_{i_j} \\
        \Rightarrow \delta Z &= \sum_{k=1}^\infty \frac{(-1)^k}{r^k} A_k \\
        \text{with }A_k &\equiv \sum_{i_1 = 3}^\infty \sum_{i_2 = i_1+1}^\infty \cdots \sum_{i_k = i_{k-1} + 1}^\infty \prod_{j=1}^k r_{i_j}~.
    \end{split}
\end{equation} 
\end{widetext}

\section{Some more relations between the coefficients and the roots of a polynomial}\label{app:moreonPOStoSOP}

In this appendix, we will list some further relations between the roots and the coefficients of a polynomial, which we crucially use in our analysis. Specifically, consider a function
\begin{equation}
    \begin{split}
        F(r) &= \sum_{n=0}^N \frac{B_n}{r^n} = \prod_{n=1}^{N} \left( 1 - \frac{R_n}{r} \right),~~~ B_0 = 1~.
    \end{split}
\end{equation}
The $R_n$'s in the second equality are the $N$ number of roots of the polynomial $F(r)$. Utilizing the expressions derived in \ref{app:POStoSOP}, one can write the coefficients $B_n$'s of the polynomial in terms of its roots $R_n$'s as
\begin{equation}
    \begin{split}
        B_n &= (-1)^n \sum_{m_1=1}^{N-(n-1)} \sum_{m_2=m_1+1}^{N-(n-2)} \cdots \sum_{m_n=m_{n-1}-1}^N \prod_{q=1}^N R_{i_q}~.
    \end{split}
\end{equation}
Our aim here will be to isolate two of the roots ($R_1$ and $R_2$), and express the combinations of sums and products of the rest of the roots ($R_3, R_4 \cdots R_N$) in terms of $R_1, R_2$, and the $B_n$'s. Casting the expressions into a more concrete form, 
\begin{widetext}
    \begin{equation}
    \begin{split}
        &\prod_{n=3}^N \left( 1 - \frac{R_n}{r} \right) \equiv 1 + \sum_{n=1}^{N-2} \frac{(-1)^n}{r^n}A_n  ~,~~~
        A_n \equiv \sum_{m_1 = 3}^{N-2-(n-1)} \sum_{m_2 = m_1+1}^{N-2-(n-2)} \cdots \sum_{m_n = m_{n-1} + 1}^{N-2} \prod_{j=1}^n R_{m_j}~.
    \end{split}
\end{equation}
Expanding the functions and comparing their coefficients, we get
\begin{equation}
    A_n = (-1)^n \sum_{m=0}^n B_m \sum_{p=0}^{n-m} R_1^p R_2^{n-m-p}~.
\end{equation}
\end{widetext}
These expressions would be useful to calculate the parameters $\alpha_j$'s while casting the effective potential into a parametrized form. There, the event horizon radii $r_+$ and $r_-$ will play the roles of $R_1$ and $R_2$, respectively, and the perturbative roots ($r_3, r_4, \cdots r_N$) will be the roots $R_3, R_4 \cdots R_N$ here.

\section{Bases of non-running TLNs}\label{app:basisTLNsnonrunning}

In this appendix, we list out the solutions to the $\gamma_n$'s in eq. (\ref{eq:recrelfull}) for the case $j \geq 2\ell+4$. For $0 \leq n \leq 2\ell$, we get $\gamma_n = 0$. 

The $\gamma_n$'s for $n> 2\ell+1$ can be expressed in terms of $\gamma_{2\ell+1}$. For the case $j\neq 2\ell+4$, for $1 \leq N \leq j-(2\ell+4)$, the recurrence relation can be solved to get
\begin{widetext}
    \begin{equation}
    \begin{split}
        \gamma_{2\ell+1+N} &= \gamma_{2\ell+1} \left[\prod_{m=1}^N \frac{(\ell+m)^2}{(2\ell+1+m)} \right] \sum_{n=0}^N \frac{1}{n! ~(N-n)!} \left(\frac{r_-}{r_+}\right)^n \frac{(\ell+n)! (\ell+N-n)!}{\ell! ~(\ell+N)!}~.
    \end{split}
\end{equation}
The last term in this sequence before the sources in the RHS of eq. (\ref{eq:recrelfull}) start contributing is given by
    \begin{equation}
    \begin{split}
        \gamma_{j-3} &= \gamma_{2\ell+1} \left[\prod_{m=1}^{j-(2\ell+4)} \frac{(\ell+m)^2}{(2\ell+1+m)} \right] \sum_{n=0}^{j-(2\ell+4)} \frac{1}{n!~ (j-2\ell-4-n)!} \left(\frac{r_-}{r_+}\right)^n \frac{(\ell+n)! (j-\ell-4-n)!}{\ell!~ (j-\ell-4))!}~.
    \end{split}
\end{equation}
For the case $j=2\ell+4$, we get  $\gamma_{j-3} = \gamma_{2\ell+1}~.$ 

For $j-2 \leq n \leq j-2 +\ell$, the sources on the RHS of eq. (\ref{eq:recrelfull}) contribute, and for $0\leq N \leq \ell$, the $\gamma_n$'s can be expressed as

    \begin{equation}
    \begin{split}
        &\gamma_{j-2+N} = \chi_N ~\gamma_{2\ell+1} \\
        &+ \sum_{m=0}^N \frac{(j-3+m)!~(j-2\ell-4+m)!}{(j-2+N)!~ (j-2\ell-3+N)!} \left[ \frac{(j-\ell-3+N)!}{(j-\ell-3+m)!} \right]^2 \phi_0^{(m)} ~ \sum_{p=0}^{\lfloor \frac{N-m}{2} \rfloor} C_{p,N,m} \left(- \frac{r_-}{r_+} \right)^p \left(1 + \frac{r_-}{r_+} \right)^{N-m-2p}  \\
        & \chi_N = \left[\prod_{m=1}^{j-(2\ell+3)+N} \frac{(\ell+m)^2}{(2\ell+1+m)} \right] \sum_{n=0}^{j-(2\ell+3)+N} \frac{1}{n!~ (j-2\ell-3+N-n)!} \left(\frac{r_-}{r_+}\right)^n \frac{(\ell+n)!~ (j-\ell-3+N-n)!}{\ell!~ (j-\ell-3+N))!} \\
        &C_{p,N,m} = \sum_{i_1=m+1}^{N-1-2(p-1)} \sum_{i_2=i_1+2}^{N-1-2(p-2)} \cdots \sum_{i_p = i_{p-1}+2}^{N-1} ~\prod_{q=1}^p \frac{(j-2+i_q)~ (j-2\ell-3+i_q)}{(j-\ell-3+i_q)~ (j-\ell-2+i_q)}~.
    \end{split}
\end{equation}
The subsequent $\gamma_n$'s for $j+\ell-1 \leq n <\infty$ can then be expressed using $1 \leq N \leq \infty$ as
    \begin{equation}
    \begin{split}
        \gamma_{j+\ell-2+N} &= \left[ \frac{(j-3+N)!}{(j-3)!} \right]^2 \left[ \frac{(j+\ell-2)!~ (j-3-\ell)!}{(j+\ell-2+N)!~ (j-3-\ell+N)!} \right] \left[ A_1 \gamma_{j-2+\ell} - \left(\frac{r_-}{r_+}\right) \left(\frac{j-3}{j-2}\right) A_2 \gamma_{j-3+\ell} \right] \\
        \text{with }A_1 &= \sum_{p=0}^{\lfloor \frac{N}{2} \rfloor} \left(-\frac{r_-}{r_+}\right)^p \left(1+\frac{r_-}{r_+}\right)^{N-2p} B_{1,p}~,\\
        A_2 &= \sum_{p=0}^{\lfloor \frac{N-1}{2} \rfloor} \left(-\frac{r_-}{r_+}\right)^p \left(1+\frac{r_-}{r_+}\right)^{N-1-2p} B_{2,p}~,\\ 
        \text{with }B_{1,p} &= \sum_{i_1=1}^{N-1-2(p-1)} \sum_{i_2=i_1+2}^{N-1-2(p-2)} \cdots \sum_{i_p = i_{p-1}+2}^{N-1} ~\prod_{q=1}^p \frac{(j+\ell-2+i_q)~ (j-\ell-3+i_q)}{(j-3+i_q)~ (j-2+i_q)} ~,~\text{     and}\\
        B_{2,p} &= \sum_{i_1=1}^{N-2-2(p-1)} \sum_{i_2=i_1+2}^{N-2-2(p-2)} \cdots \sum_{i_p = i_{p-1}+2}^{N-2} ~\prod_{q=1}^p \frac{(j+\ell-1+i_q)~ (j-\ell-2+i_q)}{(j-2+i_q)~ (j-1+i_q)}~.
    \end{split}
\end{equation}
\end{widetext}
At this stage, since we have expressed all the expansion coefficients $\gamma_n$'s in terms of $\gamma_{2\ell+1}$ and the $\phi_0^{(n)}$'s, we are now ready to argue why $\gamma_{2\ell+1}$ must be non-zero.

If one were to assume $\gamma_{2\ell+1}=0$, then it can be easily seen that for $2\ell+1 \leq n \leq j-3$, $\gamma_n = 0$. However, this leads to having non-zero values for $\gamma_n$'s with $j-2 \leq n < \infty$ owing to the non-zero contributions from the $\phi_0^{(n)}$'s. Upon performing the infinite summation over $\gamma_n (r_+/r)^n$, this leads to a potentially singular function at the horizon. In fact, this is related to the choice of arbitrary constants while directly solving for $\phi_1(r)$ from its differential equation, and the $\gamma_{2\ell+1}=0$ choice represents a bad choice of these constants that doesn't agree with the boundary condition of having a regular function at $r=r_+$, giving a logarithmic divergence there (as has been seen for different combinations of $(\ell,j)$ values.) This choice of $\gamma_{2\ell+1} = 0$ also doesn't agree with its corresponding value at the limit $r_- \to 0$ as mentioned in \cite{Katagiri:2023umb}, i.e., for the case where the radius of the inner horizon vanishes perturbatively. For all these reasons, $\gamma_{2\ell+1}$ must be non-zero. 

Following the results of \cite{Sharma:2024hlz}, since the existence of Ladder symmetry in static, spherically symmetric metrics guarantees the vanishing of Love numbers for all $\ell$ values, having a non-zero Love number for any value of $\ell$ provides a good enough counter-example for the breaking of Ladder symmetry in such spacetimes. In the following table (Table I), we have listed the values of static tidal Love numbers for several $(\ell,j)$ combinations satisfying $j \geq 2\ell+4$. 
\begin{widetext}
\begin{center}
    
\begin{table}[h!]\label{table:staticTLNnonzerol}
\begin{tabular}{ |c|c|c| } 
 \hline
 $\ell$ & $j$ & TLN \\
 \hline
  1 & 6 & $-\frac{1}{36} \alpha _6 \left(\frac{r_-^2}{r_+^2}-\frac{4 r_-}{r_+}+7\right)$ \\ 
  1 & 7 & $-\frac{1}{144} \alpha _7 \left(\frac{3 r_-^2}{r_+^2}-\frac{10 r_-}{r_+}+11\right)$ \\ 
  2 & 8 & $-\frac{1}{900} \alpha _8 \left(\frac{r_-^4}{r_+^4}-\frac{7 r_-^3}{r_+^3}+\frac{23 r_-^2}{r_+^2}-\frac{47 r_-}{r_+}+66\right)$ \\
  2 & 9 & $-\frac{\alpha _9 }{5400}\left(\frac{5 r_-^4}{r_+^4}-\frac{32 r_-^3}{r_+^3}+\frac{90 r_-^2}{r_+^2}-\frac{140 r_-}{r_+}+113\right)$ \\
  3 & 10 & $-\frac{\alpha _{10} }{98000}\left(\frac{5 r_-^6}{r_+^6}-\frac{50 r_-^5}{r_+^5}+\frac{243 r_-^4}{r_+^4}-\frac{762 r_-^3}{r_+^3}+\frac{1713 r_-^2}{r_+^2}-\frac{2892 r_-}{r_+}+3743\right)$ \\
  4 & 12 & $-\frac{\alpha _{12} }{2778300}\left(\frac{7 r_-^8}{r_+^8}-\frac{91 r_-^7}{r_+^7}+\frac{586 r_-^6}{r_+^6}-\frac{2482 r_-^5}{r_+^5}+\frac{7699 r_-^4}{r_+^4}-\frac{18403 r_-^3}{r_+^3}+\frac{34810 r_-^2}{r_+^2}-\frac{52882 r_-}{r_+}+65056\right)$ \\
  4 & 14 & $-\frac{\alpha _{14} }{30561300}\left(\frac{63 r_-^8}{r_+^8}-\frac{756 r_-^7}{r_+^7}+\frac{4340 r_-^6}{r_+^6}-\frac{15729 r_-^5}{r_+^5}+\frac{39747 r_-^4}{r_+^4}-\frac{72566 r_-^3}{r_+^3}+\frac{95178 r_-^2}{r_+^2}-\frac{84543 r_-}{r_+}+41126\right)$ \\
 \hline
\end{tabular}
\caption{Values of static Tidal Love numbers for different $\ell$ and $j$ combinations}
\end{table}
\end{center}
\end{widetext}

\section{Parametrized Love numbers for KRZ class of Metric}\label{app:KRZTLN0}

In this appendix, we derive the tidal response for the Ladder-symmetric class of stationary rotating KRZ metrics, and show that TLN vanishes for them.

With $V_{\ell} = \frac{\ell(\ell+1)}{r^2} + \frac{\partial_r f_0}{r} - \frac{a^2 m^2}{r^4 f_0}$, the Klein-Gordon equation for a massless static scalar field becomes
\begin{equation}\label{groundstate}
    \frac{d}{dr}\left[\bar{f} \frac{d\phi_0(r)}{dr}\right] - V_{\ell} \phi_0(r) = 0~.
\end{equation}

\begin{widetext}
Solving, we get
\begin{equation}
    \phi_0(r) = c_1 r P_\ell^{\frac{2 i a}{r_+-r_-}}\left(\frac{2 r}{r_+-r_-}-\frac{r_++r_-}{r_+-r_-}\right)+c_2 r Q_\ell^{\frac{2 i a}{r_+-r_-}}\left(\frac{2 r}{r_+-r_-}-\frac{r_++r_-}{r_+-r_-}\right)~.
\end{equation}

We rewrite this solution in terms of a regularized hypergeometric function as
\begin{equation}
\begin{split}
\phi_0(r) = & \frac{1}{4} i r \, 
e^{\tfrac{\pi a}{r_--r_+}}
\left(\frac{r-r_+}{r-r_-}\right)^{\tfrac{i a}{r_--r_+}}
\Bigg( 
 \frac{1}{\Gamma\!\left(1+\ell+\tfrac{2 i a}{r_--r_+}\right)}
    2 \pi c_2 
    \left(\frac{r-r_+}{r-r_-}\right)^{\tfrac{2 i a}{r_+-r_-}}
    \text{csch}\!\left(\tfrac{2 \pi a}{r_+-r_-}\right)
    \Gamma\!\left(1+\ell+\tfrac{2 i a}{r_+-r_-}\right) \, \\
    & _2\tilde{F}_1\!\left(-\ell,\ell+1;1+\tfrac{2 i a}{r_+-r_-};\tfrac{r_+-r}{r_+-r_-}\right) - e^{\tfrac{2 \pi a}{r_+-r_-}} \tanh\!\left(\tfrac{\pi a}{r_+-r_-}\right) \Big( \pi c_2 \coth^2\!\left(\tfrac{\pi a}{r_+-r_-}\right) + 4 i c_1 \coth\!\left(\tfrac{\pi a}{r_+-r_-}\right)  \\
    & \quad + \pi c_2 \Big) \, _2\tilde{F}_1\!\left(-\ell,\ell+1;1+\tfrac{2 i a}{r_--r_+};\tfrac{r_+-r}{r_+-r_-}\right)
\Bigg)~.
\end{split}
\end{equation}
At this stage, we can apply the two boundary conditions to fix the constants: we demand only an ingoing mode at the horizon, which fixes one of the constants, $c_2$, and $c_1$ is such that the coefficient of $\left(\frac{r}{r_+}\right)^{\ell+1}$ is unity. Therefore, we must choose the two constants as
\begin{equation}
    c_1 = \frac{1}{2} i \pi  c_2 \coth \left(\frac{2 \pi  a}{r_+-r_-}\right)~,
\end{equation}
\begin{equation}
    c_2 =\frac{i \Gamma (\ell+1) e^{\frac{3 \pi  a}{r_+-r_-}} \left(e^{\frac{4 \pi  a}{r_--r_+}}-1\right) \left(1-\frac{r_-}{r_+}\right)^\ell \Gamma \left(1+\ell+\frac{2 i a}{r_--r_+}\right)}{\pi  r_+ \Gamma (2\ell+1)}~.
\end{equation}
Substituting this back in our expression of $\phi_0(r)$, we get the final zeroth order horizon regular solution,
\begin{equation}
    \phi_0(r)= \frac{r \Gamma (\ell+1) \left(1-\frac{r_-}{r_+}\right)^\ell \left(\frac{r-r_+}{r-r_-}\right)^{\frac{i a}{r_+-r_-}} \Gamma \left(1+\ell+\frac{2 i a}{r_+-r_-}\right) \, _2\tilde{F}_1\left(-\ell,\ell+1;1+\frac{2 i a}{r_+-r_-};\frac{r_+-r}{r_+-r_-}\right)}{r_+ \Gamma (2\ell+1)}~.
\end{equation}
To obtain the love number at this order, we expand the solution at large r and compare it with equation $(17)$ of \cite{Katagiri:2023umb}. We get
\begin{equation}
\begin{split}
    \phi_0(r)\big|_{r \gg r_+} 
    &= \left(\frac{r}{r_+}\right)^{\ell+1}
       \left[1+\mathcal{O}\!\left(\frac{r_+}{r}\right)\right] \\
    &\quad + 
    \left(
      \frac{(r_+-r_-)^{2 \ell+1}\Gamma (-2 \ell-1) \Gamma (\ell+1) \Gamma \left(1+\ell+\frac{2 i a}{r_+-r_-}\right)}{r_+^{2 \ell+1} \Gamma (-\ell) \Gamma (2 \ell+1) \Gamma \left(-\ell+\frac{2 i a}{r_+-r_-}\right)}
    \right)
    \left(\frac{r_+}{r}\right)^{\ell}
    \left[1+\mathcal{O}\!\left(\frac{r_+}{r}\right)\right]~.
\end{split}
\end{equation}
From the above expression, we can read off the TLN as
\begin{equation}
    \kappa_{(0)}^{KRZ} =
      \frac{(r_+-r_-)^{2 \ell+1}\Gamma (-2 \ell-1) \Gamma (\ell+1) \Gamma \left(1+\ell+\frac{2 i a}{r_+-r_-}\right)}{r_+^{2 \ell+1} \Gamma (-\ell) \Gamma (2 \ell+1) \Gamma \left(-\ell+\frac{2 i a}{r_+-r_-}\right)}~,
\end{equation}
which can be simplified further to get
\begin{equation}
    \kappa_{(0)}^{KRZ} = -\frac{i a}{r_+ - r_-}\frac{{\Gamma(\ell+1)}^2}{\Gamma(2\ell+1)\Gamma(2\ell+2)}\left(1-\frac{r_-}{r_+}\right)^{2\ell+1} \prod_{j=1}^{\ell} \left(j^2 + \frac{4 a^2}{(r_+ - r_-)^2}\right)~.
\end{equation}
\end{widetext}
The above expression for the rotating KRZ metric is purely imaginary and therefore, it is regarded as tidal dissipation number. Its real part which is the static TLN, vanishes for all $\ell$ and $m$.

\bibliography{ladsym}

\end{document}